\documentclass[12pt]{article}
\usepackage{graphics,amsmath }
\begin{document}

\title{ \bf Microscopic Study Reveals the Singular Origins of Growth.}
\author{Gur Yaari$^{1,2}$, Andrzej Nowak $^{3}$,  Kamil Rakocy $^{3}$  \&  Sorin Solomon$^{2,1}$}
%

%
\maketitle
\bigskip$^1$ {\small Institute for Scientific Interchange, via S. Severo 65, I-10113 Turin, Italy}\\
\bigskip$^2$ {\small Racah Institute of Physics, Hebrew University, IL-91904 Jerusalem, Israel}\\
\bigskip$^3$ {\small Center for Complex Systems, Institute for Social Studies, University of Warsaw 00 - 183, Poland}\\
\abstract{
Anderson \cite{anderson1972} proposed the concept of complexity in order to describe the emergence and growth of macroscopic collective patterns out of the simple interactions of many microscopic agents. In the physical sciences this paradigm was implemented systematically and confirmed repeatedly by successful confrontation with reality. In the social sciences however, the possibilities to stage experiments to validate it are limited.  During the 90's a series of dramatic political and economic events have provided the opportunity to do so. We exploit the resulting empirical evidence to validate a simple agent based alternative to the classical logistic dynamics. The post-liberalization empirical data from Poland confirm the theoretical prediction that the dynamics is dominated by singular rare events which insure the resilience and adaptability of the system. We have shown that growth is led by few singular "growth centers" (Figure \ref{fig1}), that initially developed at a tremendous rate (Figure\ref{fig3}), followed by a diffusion process to the rest of the country and leading to a positive growth rate uniform across the counties. In addition to the interdisciplinary unifying potential of our generic formal approach, the present work reveals the strong causal ties between the "softer" social conditions and their "hard" economic consequences. 
} 
\section{Introduction}
\label{intro}
The essence of economic growth and growth in general was captured by T.R. Malthus\cite{Malthus1798} and P.F. Verhulst\cite{Verhulst1838} as early as the 1800's: multiplication and competition.
Malthus was the first to write (1798) a differential equation describing the dynamics of a population of individuals proliferating at a fixed growth rate $a$: 
\begin{equation}
  \frac{d W (t)}{d t} = a \cdot W (t) \label{eqMAL}
\end{equation}

The linear term $a \cdot W (t)$ can represent a very wide range of "auto-catalytic" phenomena in various fields:  proliferation in biology, returns in economics , or proselytizing in politics. Malthus believed that the (exponential) solution $\sim e^{(a\cdot t)}$    of his equation would cause a humanitarian “catastrophe”. However, Verhulst {\cite{Verhulst1838}} introduced (in 1838) a nonlinear interaction term $-b\cdot W^2$ (that may represent confrontation in biology, competition in economics, and limited constituency in politics) 
\begin{equation}
  \frac{d W (t)}{d t} = a  \cdot W (t) - b\cdot W^2 (t) \label{eqVER}
\end{equation}

By including this term the solution approaches smoothly a constant value $W \longrightarrow \frac{a}{b}$ instead of the catastrophic divergence.  
During the following two centuries, this "logistic dynamics" was recognized by leading scientists as fundamental in various fields from biology (Volterra{\cite{Volterra1931}}) to "the everyday world of politics and economics" (Robert May{\cite{May1976}}). 

Ronald Ross was awarded the Nobel prize\cite{Ross1911} for applying those ideas to the spread of malaria in humans and mosquitos. The model was expressed by Lotka{\cite{Lotka1923}} in terms of a coupled system of two equations generalizing \eqref{eqVER}: 

\begin{equation} \label{eqLOT}
\begin{array}{cc}
\frac{d w_1 (t)}{d t} =a_1\cdot w_1 (t) +a_{12}\cdot w_2 (t) -a_{112}\cdot w_1 (t)\cdot w_2 (t)
 \\
\frac{d w_2 (t)}{d t}=a_2\cdot w_2 (t) +a_{21}\cdot w_1 (t) -a_{212}\cdot w_1 (t)\cdot w_2 (t)
\end{array}
\end{equation}
Lotka was able to study the stability properties of such equations and their numerical treatment and to predict the ratios between the infected mosquitoes and the infected humans. By the same time, Vito Volterra advocated independently the use of equations in biology and social sciences.\cite{Volterra1931} In particular, Volterra re-deduced the logistic curve by reducing the Verhulst equation \eqref{eqVER} to a variational principle (maximizing a function that he named "quantity of life"\cite{Lotka1923}). 
The extension of \eqref{eqVER} to spatial distributed systems in terms of partial differential equations was first formulated by R.A. Fisher\cite{Fisher1937}:
\begin{equation}\label{eqFIS}
  \frac{\partial W (\overrightarrow{x},t)}{\partial t} = a \cdot W (\overrightarrow{x},t) - b\cdot W^2 (\overrightarrow{x},t) +D\cdot\nabla^2 W (\overrightarrow{x},t)
\end{equation}
in the context of the propagation ("Fisher wave") of a mutant superior gene in a population. The mathematical study of {\eqref{eqFIS} was taken up by mathematicians{\cite{Kesten1980}} and followed up by a large literature in physics journals.\cite{BenAvraham_Havlin2000,Grassberger1982,Janssen1981,Cardy_Tauber1996}

A further generalization (from two populations and two equations to an arbitrary number of populations and equations) of the logistic system (3) was proposed by Eigen\cite{Eigen1971} and Eigen and Schuster\cite{Eigen_Schuster1979} in the context of Darwinian selection and evolution in prebiotic environments. 
They considered quasi-species of auto-catalytic (self reproducing RNA) molecules which can undergo mutations. The model considers auto-catalytic sequences $i$ self-replicating at various rates $a_i$ and undergoing mutations to other sequences $j$ at rates $a_{ij}$ . 
\begin{equation}\label{eqEIG}
 \begin{array}{cc}  
\frac{d W_i (t)}{d t} = a_i \cdot W_i (t) + {\sum_{j = 1}^N} a_{ij} \cdot W_j (t)\\
- {\sum_{j = 1}^N} a_{ji} \cdot W_i (t)- b (  {\overrightarrow{W (t)}, t)}) \cdot W_i (t)
                          \end{array}
\end{equation}

The term $b({\overrightarrow{W (t)}, t)}$ represents generically the interaction with the environment (in the specific case of ref \cite{Eigen_Schuster1979}  the result of replenishing and steering the container continuously). 
Montroll proposed the extension of the logistic framework to social sciences: "almost all the social phenomena, except in their relatively brief abnormal times obey the logistic growth."{\cite{Montroll1978}}. 

In economics, the ecology-market analogy was recognized quite early (e.g. \cite{Schumpeter1934} ) and it was developed in parallel with the other, physical analogies. This lead eventually to a wide literature in evolutionary economics (e.g.\cite{Nelson_Winter1982,Ebeling_Feistel1982,Jimenez-Montano_Ebeling1980}). Within the extension of \eqref{eqEIG} to economics, the $a_i$ 's may represent individual or corporate capital growth rates, GDP growth rates etc. while the $a_{ij}$'s terms may represent the effects of trade, social security, some form of mutual help or other economic mechanisms resulting in wealth transfer (e.g. taxes / subsidies). 
One of the finance applications{\cite{Marsili_ea1998}} interprets $i$ as the equities within a personal portfolio, $a_i(t)$ as the rate of growth of the equity $i$ (at time $t$) and $a_{ij}$ as the periodic redistribution of capital between the equities by the owner of the portfolio (in order to optimize it). 
Stochastic generalizations of the logistic/ Lotka-Volterra equations were studied also in a large body of mathematical literature (e.g. \cite{Kesten1980}), and in order to get meaningful results out of the model, one has to introduce the noise in a proper way that will stand for it's effect in real-life systems. 
In conclusion, generic logistic ideas hinted by \eqref{eqVER} arose for the last century in an extremely wide-ranging set of applications. For each discipline, subject and system, the variables of the model had to be interpreted in terms of the empirical observables and adapted to the relevant range of parameters and initial conditions. Once the parameters are specified, the generic framework \eqref{eqEIG} (plus noise) becomes a well defined model for a specific system. Then, one can derive from it precise predictions and confront them with the data.

In the present work we analyzed a system of the generic type \eqref{eqEIG} as applied to the conditions and constraints of economic growth. In this framework we were able to deduce rigorously a series of nontrivial and sometimes unexpected or even superficially contradictory predictions. All those predictions turned out to be confirmed in great detail by the empirical data as seen below. Those predictions are fundamentally different form the global behavior that would have been obtained by applying naively the global equation \eqref{eqVER} to the system as a whole. As such, they are expressions of the importance of the spatio-temporal inhomogeneity in multi-agent complex models. The results are also fundamentally different from a spatially continuous extended system: indeed as shown in \cite{Shnerb_ea2000,Shnerb_ea2001} a system of partial differential equations would lead to predictions in complete contradiction with the actual behavior of the spatially discrete system: the continuity assumption blurs the singular character of the growth centers which it is the dominant feature of the present system.

\section{The generic theoretical framework applied to the Post Liberalization Polish Economy.}
\label{sec1}

The historical occasion to validate the model arose following the political "big bang" of the Soviet Bloc in the 1990's. With this occasion one was able to empirically observe the unfolding of the capitalist \cite{Rosser_Rosser2004} dynamics represented by \eqref{eqEIG} from the well defined initial conditions of the socialist, centrally planned economy. Along to generic features common to all countries, this liberalization transition has led to various features that are country dependent. In some countries, (e.g. Poland), not long after the transition, the economy started to grow at a fast rate soon surpassing the level of its economy under socialism. In others (e.g. Ukraine) the transition resulted in a long lasting economic crisis that only recently has started to show observable improvement. The system \eqref{eqEIG} is capable to explain within one generic theoretical framework both the universal facts and the various particular features. 

We present below a very detailed empirical check of the model by applying it to detailed economic data collected in Poland after the liberalization. 
In this application, the system \eqref{eqEIG} consists of 2945 equations. The equations are indexed by an index $i$. representing one of the 2945 counties composing Poland. Each equation represents the evolution of the economic activity $W_i$ of the county $i$. As the main proxy for the density of the economic activity we used the number of enterprises per capita. The endogenous growth rate is denoted by $a_i$ and may depend on various factors such as social capital, availability of natural resources or infrastructure in $i$. One of the most surprising and widely relevant findings of our study is that the purely economic quantities $a_i$ turned out to be strongly dependent on social factors. One of the social factors impinging most on $a_i$ turned out to be the education level (Figures:\ref{fig1},\ref{fig2}). The counties with the highest education level will be indicated by the index $i = max$ , such that $a_{max}$ denote the largest endogenous growth rates after liberalization. 
In general, the $a_i$'s include capital loss, consumption and depreciation and thus they  may be negative. In fact, in the case of post-liberalization Poland, the majority of the $a_i$'s as well as their average $<a_i>_i$ turned out to be negative (we will see that in spite of this, as predicted by theory, the average economic activity $< W_i>_i$  increases asymptotically).
The terms $a_{ij}\cdot W_j$ in \eqref{eqEIG} should then be interpreted as the flow (diffusion/transfer) of economic activity from county $j$ to county $i$ . We assume no part of the country is completely isolated of the rest, i.e. the matrix $a_{ij}$ is not block diagonal (in particular we assume that the eigenvector $\hat{n_i}$ with the highest eigenvalue has non-vanishing components for all the $i$'s). The transfer rates $a_{ij}$ are larger for neighboring $i$ and $j$ counties due to transportation costs,\cite{Thunen1826,Weber1909} local positive feedback loops\cite{Arthur1994}and social influence\cite{Nowak_ea2001,Nowak_ea1994}. The term $b({\overrightarrow{W (t)}, t)}$ represents the arbitrary nonlinear interactions among the counties. Moreover $b(\overrightarrow{W (t)}, t)$ is an arbitrary function of  time and the state of the whole system $\overrightarrow{W (t)}$, thus it can express the global state of the economy (boom, crash, etc.) under the influence of the national and international social, cultural, economic and political environment (however in the present paper we are mainly interested in the effect of the endogenous factors after liberalization).

Within the present particular application, the generic properties of \eqref{eqEIG} concretize in the following predictions:
\renewcommand{\labelenumi}{\Alph{enumi})}
\renewcommand{\labelenumii}{\alph{enumii})}
\renewcommand{\theenumi}{\Alph{enumi})}
\begin{enumerate}
 \item Initially the various counties develop at widely different rates as opposed to the partial differential models' results where the growth rate has a smooth spatial variation.\label{A}
\item At this stage all the counties have comparable GDP's and most of the counties have negative growth rates, so the total GDP is decaying. \label{B}
\item Following liberalization the distribution of the counties' economic activity widens significantly leading to order of magnitudes differences between the economic activity per capita $W_{max}$ of the fast developing counties (with large positive $a_i=a_{max}$) and the rest of the  counties.\label{C}
\item Consequently in the later stages, the centers with the highest endogenous growth rate $a_i=a_{max}$  constitute a larger part $W_{max}$ of the total GDP and lead to its' growth. \label{D}
\item The predictions \ref{B} and \ref{D} amount to the "J-curve" \cite{Brada_King1992} effect: the total GDP decreases first and then increases. This is at stark variance to the usual logistic curve that is monotonic.\label{E}
\item As time passes, the positive growth rate regime diffuses to an ever increasing distance from the growth centers. \label{F}
\item Eventually the entire country reaches the asymptotic stage where the growth rates of all counties $\frac{dW_i}{dt}$ have a common positive value.
The intuitive explanation of this analytic result is that the transfer of economic activity from the rich regions dominates eventually the local endogenous growth of the slow regions.\label{G}
\item In spite of this, the spatial distribution of the economic activity per capita itself continues to be concentrated in limited regions around the growth centers initially revealed at \ref{A}) and \ref{D}).\label{H}
\item Thus, even in the asymptotic regime, the differences between counties' economic activities continue to increase exponentially. However since in this stage the whole country share the same growth rate, the \textbf{\textit{ratio}} between the economic activity of any two counties remains constant.\label{I}

\end{enumerate}

The theoretical predictions \ref{A}-\ref{I} were confirmed by the empirical data:  
Immediately after the liberalization, the Polish counties behaved in dramatically different ways (leading to ever increasing inequality: prediction \ref{C} and Figure\ref{fig5}: while most of the counties' economies plunged by factors of two, a few counties tripled their economic activity (Predictions \ref{A} and Figure\ref{fig3}. During the preceding (socialist) regime, these active counties were not particularly well developed, so to begin with, their relative weight in the global GDP was too small to avert the fast decay (Prediction \ref{B} and \ref{E} and Figure\ref{fig4}). Later, following their dramatic growth, the fast developing counties became the economic force, driving up the GDP (Prediction \ref{D},\ref{E} and Figure\ref{fig3}). The influence of the growing centers expands with time to wider regions (Prediction \ref{F})and Figure\ref{fig1}). Consequently the entire country eventually reaches a spatially uniform growth regime (Prediction \ref{G} and Figure\ref{fig3}). Still, even in the spatially uniform growth rate regime, the very wide differences in the economic activity per capita persisted (Prediction \ref{H} and \ref{I} and  Figures \ref{fig2},\ref{fig3} and \ref{fig5}). 
In conclusion the theoretical analysis predicted a coherent (if unexpected) narrative of the post-liberalization economy which was completely confirmed by the data:
An interesting unexpected observation emerging from our analysis of the data is the role of education in the economic development: The analysis of the empirical data through the glasses of the model \eqref{eqEIG} shows that the counties' economic activity \textit{after} the liberalization (Figure\ref{fig2}B) has a strong correlation with the education distribution (generated by the socialist regime) \textit{before} liberalization (Figures \ref{fig2}C and \ref{fig6}). Ironically, education level was irrelevant for the economy distribution before liberalization (Figures \ref{fig2}A and \ref{fig6}).
To summary, the liberalization of the economies of the Soviet bloc following its disintegration\cite{Aghion_Blanchard1994,Levy_ea2000} was used here as a laboratory to study economic growth. Using high space-time resolution data from Poland we show that the liberalization transition follows a "microscopically" discretized version of the classical logistic dynamics.\cite{Verhulst1838,Lotka1923} 
\section{Discussion and Conclusions}

The studies of growth dynamics that use data and models at a coarse level of aggregation (e.g. representing them in terms of smooth spatio-temporal densities governed by (/partial) differential equation) have difficulties to provide a generic explanation of growth dynamics.  The present “microscopic representation” \cite{Levy_ea2000} approach allows us to connect complex macroscopic collective trends to their simple local causes in a wide range of applications. In particular this approach uncovers the role of singular localized auto-catalytic growth centers. 
In the present paper we presented a detailed confrontation between the theoretical predictions and the empirical data. 

The liberalization of the economies of the Soviet bloc following its disintegration\cite{Aghion_Blanchard1994,Rosser_Rosser1997} was used here as a laboratory to study economic growth. Using high space-time resolution data from Poland we show that the liberalization transition follows a “microscopically” discretized version of the classical logistic dynamics.\cite{Verhulst1838} 
We find that the application of the classical logistic dynamics ideas to growth is warranted if (and only if) one takes appropriately into account the dynamic implications of 'microscopic' inhomogeneity, randomness and especially the role of the singular, localized and rare events. 
In light of the empirical validation for the present theoretical framework we are in the process of applying it to other economic experiments and to different aggregation levels (sub-county and international).
We propose to extend this methodology in further possible applications such as population biology,\cite{Louzoun_ea2003} immunology,\cite{Louzoun_ea2001a} markets,\cite{Louzoun_ea2003a} and Internet.\cite{Goldenberg_ea2005}

\appendix{\textbf{Acknowledgements}}
We acknowledge very instructive discussions with Giulio Bottazzi, Giovanni Dosi, Mauro Gallegatti, Moshe Levy, Yoram Louzoun, Thomas Lux, David Bree, Diana Mangalagiu, Barklay Roser , Nadav Shnerb, Dietrich Stauffer, Gerard Weisbuch, Giancarlo Mosetti, Joseph Wakeling and Amos Ilan. David Bree help was crucial in expressing a few of the ideas. The present research was partially supported by the STREPs CO3 and DAPHNet of EC FP6.

\appendix{\textbf{Appendix A. Methods:}}
The various regimes of the system \eqref{eqEIG} determine the theoretical predictions A-I at various stages in the post-liberalization process. Here we give some explanations of how those predictions are extracted:
\renewcommand{\labelenumi}{\Alph{enumi})}
\renewcommand{\labelenumii}{\alph{enumii})}

\begin{enumerate}

\item  Under the assumption that the initial $W_i(0)$  's are distributed randomly uniform, the diffusion terms ${\sum_{j = 1}^N} a_{ij}\cdot W_j (t)$ and  $ {\sum_{j = 1}^N} a_{ji}\cdot W_i (t)$compensate one another and the various counties evolve initially with the different effective growth rates: 
\begin{equation}
 \frac{d W_i (t)}{d t} \sim [a_i  -b (
  {\overrightarrow{W (t)}, t)})] \cdot W_i (t)  	
\end{equation}

Thus, with the exception of the few counties with the largest positive $a_i$'s most of the counties GDP's decay initially.	

\item  Given the assumption that the initial   $W_i(0)$ 's are random and in particular uncorrelated to the $a_i$'s the initial variation of the national average of enterprises per capita which defines as:
\begin{equation}
  W ( t ) = {\sum_{i = 1}^N} \frac{W_i ( t )}{N}
\end{equation}
factorizes: 
\begin{equation}\label{eq9}
 \frac{d W ( t )}{d t} \sim [<a_i>_i  -b (
  {\overrightarrow{W (t)}, t)})] \cdot W ( t )  	
\end{equation}
Since the large majority of the counties have negative growth rate and  $<a_i>_i<0$   one has immediately after $t=0$ a negative average growth rate:  
\begin{equation}
 \frac{d W ( t )}{d t} <0
\end{equation}

\item  If one assumes that the $a_i$ 's are stochastically independent, the economic activities in different counties $W_i$ develop following independent growth/ decay rates. This leads to the prediction of a widening histogram of $W_i$ 's with an approximate log-normal shape. As the system enters in the later phases (\ref{G}-\ref{I} below) the growth rates of various counties become equal \eqref{eqEIG2} and the relative distribution stabilizes. 

\item  As time passes, $W_i(t)$ becomes increasingly correlated to $a_i$: the $W_i$ 's corresponding to the largest positive  $a_i = a_{max}$ becomes the largest $W_i(t)$' s. Thus the factorization in \eqref{eq9} ceases to hold and the variation of the average $\frac{d W ( t\gg 0 )}{d t} $ is dominated by the largest positive terms 

\begin{equation}\label{eq11}
 \frac{d W ( t )}{d t} \sim  [a_{max} -b (
 {\overrightarrow{W (t)}, t)}) ]\cdot W_{max}(t) > 0
\end{equation}
\item  From \ref{C} and \ref{D} (and assuming the external terms $b ({\overrightarrow{W (t)}, t)}$ are not dominant) the system generically passes from initial decay \eqref{eq10} to eventual growth \eqref{eq11}.
This is the “J-curve” shape revealed in all the post-liberalization economies. 

\item  See \ref{G}

\item  If the matrix $A=\parallel a_{ij}\parallel$ is not block diagonal, after some time, all the counties will share a common (though in general - due to $b ({\overrightarrow{W (t)}, t)}$ - time dependent) growth rate:  
After regrouping the terms proportional to $W_i$ in \eqref{eqEIG} one obtains:
\begin{equation}\label{equ2}
  \frac{d W_i ( t )}{d t} = ( a_i - {\sum_{j = 1}^N} a_{j i} + b ({\overrightarrow{W (t)}, t)}) \cdot W_i ( t ) + {\sum_{j = 1}^N} a_{i j} \cdot
  W_j ( t ) 
\end{equation}
then, by summing \eqref{equ2} over $i$, one gets:
\begin{equation}\label{eq6} 
  \frac{d W ( t )}{{dt}} = b ({\overrightarrow{W (t)}, t)} \cdot W ( t ) +
  {\sum_{i = 1}^N} a_i \cdot \frac{W_i ( t )}{N}
\end{equation}
The dynamics of $W ( t )$ and of $W_i ( t )$ can be very complicated, mainly
due to the time dependent function $b ({\overrightarrow{W (t)}, t)}$. Nevertheless if one is
interested in the proportional wealth of each group:
\begin{equation}
  X_i ( t ) \equiv \frac{W_i ( t )}{W ( t )} \label{eq7}
\end{equation}
then, by writing the chain differential rule:
\begin{equation}
  \frac{d X_i}{{dt}} = \frac{1}{W} \frac{d W_i}{d t} - W_i 
  \frac{1}{W^2} \frac{d W}{d t} = \frac{1}{W} \frac{d W_i}{d t} - X_i 
  \frac{1}{W} \frac{d W}{d t}
\end{equation}
and plugging in equations \eqref{equ2}, \eqref{eq6} and \eqref{eq7} one gets:
\begin{equation}
  \frac{d X_i}{d t} = ( a_i - {\sum_{j = 1}^N} \frac{a_j X_j ( t
  )}{N} - {\sum_{j = 1}^N} a_{j i} )  X_i ( t ) +
  {\sum_{j = 1}^N} a_{i j}  X_j ( t ) \label{eq10}
\end{equation}
(notice how the $b ({\overrightarrow{W (t)}, t)}$ had been canceled)
A steady state distribution is reached iff $\frac{{dX}_i}{{dt}}
{=} 0$, which implies:
\begin{equation}\label{equation12}
  ( a_i - {\sum_{j = 1}^N} a_{j i} )  X_i ( t ) +
  {\sum_{j = 1}^N} a_{i j} X_j ( t ) = \frac{1}{N}  (
  {\sum_{j = 1}^N} a_j X_j ( t ) ) X_i ( t ) 
\end{equation}
Which is equivalent with the eigenvalue problem:
\begin{equation}\label{eq13}
  ( a_i - {\sum_{j = 1}^N} a_{j i} )  X_i ( t ) +
  {\sum_{j = 1}^N} a_{{ij}}  X_j ( t ) = \Lambda X_i
  ( t ) 
\end{equation}
Once \eqref{eq13} is valid $\Lambda$ could be found by averaging over
all the equations and using :
\begin{equation}\label{eq14}
  < X ( t ) > = \frac{1}{N} {\sum_{j = 1}^N} \frac{W_j ( t )}{W ( t
  )} = 1 
\end{equation}
to be consistent with equation \eqref{equation12} :
\begin{equation} \label{eq15}
\begin{array}{cc}
 \frac{1}{N} {\sum_{i = 1}^N} \{ ( a_i - {\sum_{j =
  1}^N} a_{j i} ) X_i ( t ) + {\sum_{j = 1}^N} a_{i j} X_j ( t )
  \} =\\
= \frac{1}{N} ( {\sum_{j = 1}^N} a_j X_j ( t ) ) = \Lambda
\end{array}
\end{equation}
once the steady state of equation \eqref{eq10} is reached, the actual time
evolution of the different counties is expressed (according to \eqref{equ2},
\eqref{eq7}, \eqref{eq13}) as:
\begin{equation} \label{eq18}
\begin{array}{cc}
 \frac{d W_i ( t )}{{dt}} = ( b ( \overrightarrow{W ( t )}, t ) + a_i ) W_i ( t ) + ( \Lambda - a_i )  W_i ( t ) =\\
= ( \Lambda + b ( \overrightarrow{W ( t )}, t ) ) \cdot W_i ( t ) 
\end{array}
\end{equation}
What can we say about the solutions of \eqref{eq13}? By moving $( a_i -
{\sum_{j = 1}^N} a_{j i} ) \cdot X_i ( t )$ to the rhs and dividing
by $( \Lambda - a_i + {\sum_{j = 1}^N} a_{j i} )$ we can see:
\begin{equation}
  X_i ( t ) = \frac{{\sum_{j = 1}^N} a_{i j} \cdot X_j ( t )}{(
  \Lambda - a_i + {\sum_{j = 1}^N} a_{j i} )} \label{eq17}
\end{equation}
Now, if we assume that all the $a_{i j}$'s are positive and admit only
positive $X_i$ values then the eigenvalues allowed are when $\Lambda > ( a_i -
{\sum_{j = 1}^N} a_{j i} ) \equiv \tilde{a}_i$ for all $i$'s. In
particular :
\begin{equation}
 \label{eq16} \Lambda > \tilde{a}_{\max}
\end{equation}
where $max$ refers to the index of the largest $X_i$ (and $\tilde{a}_i$).
But in fact one can easily see that there is only one such solution (the other
$\Lambda$ eigenvalues are placed each in one of the real axis intervals
delimited by two successive $\tilde{a}_j$ values, and from equation
\eqref{eq16}, we have only one solution). 
Once we have reached a steady state for the $X_i$'s one could easily see that
the ratio's between the different $W_i ( t )$'s have the same values as the
ratios between the $X_i$'s (recall equation \eqref{eq7}) This implies that after 
long enough time (after reaching the steady state) the different counties grow 
with the same rate. 
 \item  As $W_{max}$ become dominant, the growth of the other $W_i$ 's is proportional mainly to $a_{i max}\cdot W_{max}$  . Since the transfer terms $a_{i max}$ are largest for counties $i$ neighboring the growth centers $j=max$ , the spatial distribution of the economic growth diffuses in the first stage to the immediate neighborhood of the growth centers. As time passes the growth regime diffuses to an ever-widening area. In a matter of fact, the local property of the matrix  $A=\parallel a_{ij}\parallel$, connects the system \eqref{eqEIG} to a diffusing reactant agent based model (the "AB model"\cite{Shnerb_ea2000,Shnerb_ea2001,Louzoun_ea2003a,Louzoun_ea2003,Louzoun_ea2007}). The spatial profile of the economic activity has been shown in that context to decay exponentially, leading to regions of high activity localized around the original growth centers.	
\item  The uniformization of the growth rates $\frac{d W_i (t)}{d t} $  to $\nu_{max}(t)$ in \eqref{equation12} does not lead to the convergence of counties GDP's  $W_i$ but rather to a stable ratio between them. Indeed, the values of the $W_i$'s are given by the components of the highest eigenvector $\hat{n_i}(t)$ of $M_{ij} (t)$ times a term that grows as $e^{\int \nu_{max}(t)dt}$, when $M_{ij} (t)$ is taken from the rhs of \eqref{eqEIG}, i.e: 
\begin{equation}\label{eqEIG2}
\begin{array}{cc}
\frac{dW_i(t)}{dt}=\sum_j^NM_{ij} (t)\cdot W_j(t) \equiv \\ \equiv \sum_j^N[\delta_{ij}(a_i-\sum_k^Na_{ki}-b(\overrightarrow{W(t)},t))+a_{ij}]\cdot W_j(t)\\=\nu_{max}\cdot W_j(t)
\end{array}
\end{equation}
  Thus, differences between the counties GDP's diverge asymptotically as $e^{\int \nu_{max}(t)dt}$..
\end{enumerate}
%
\begin{figure*}
\resizebox{1\columnwidth}{!}{
  \includegraphics{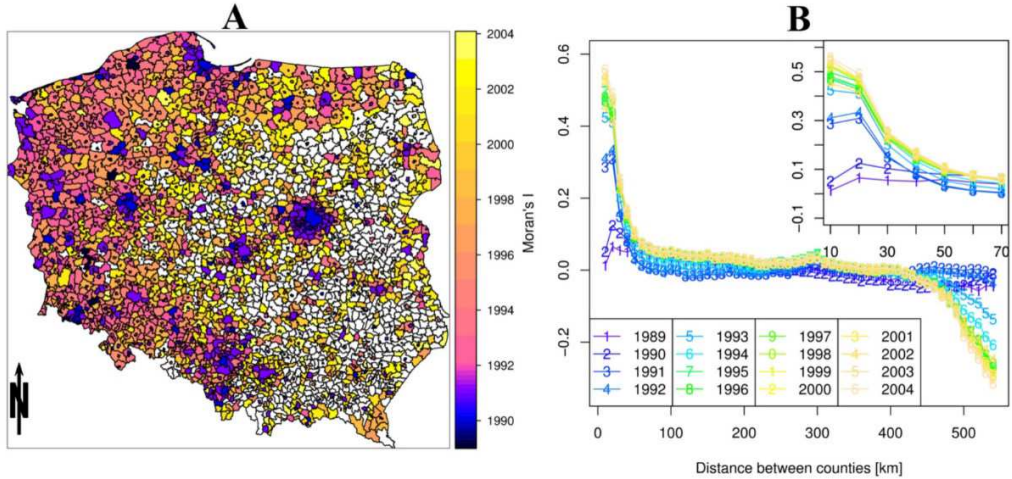}
}
\caption{\textbf{The spread of economic growth from the singular growth centers to the rest of the country.} \textbf{ \ref{fig1}A}: Immediately after liberalization, the singular centers colored in dark blue tripled their economic activity and exceeded the threshold of 0.01 enterprises per capita. The rest of the colors visualize the spread of economic growth by marking the year in which each specific county has passed the 0.01 threshold. One sees that the spread process has a characteristic pattern: first the appearance of an isolated singular growth center followed by the spread of the growth to its immediate neighbors. Eventually most of the country is "conquered" by progress. \textbf{ \ref{fig1}B}: The statistical analysis of the correlations between the economic activity of counties as a function of their distance confirms the visual intuitions expressed by \textbf{ \ref{fig1}A}. Indeed, one sees that at the beginning of the process the different counties were not correlated while towards the end of the process the influence of the economic activity of one county propagated to quite far away counties.The spatial correlation is calculated with the aid of Moran's index which is defined as: $ I \equiv \frac{N {\sum_i} {\sum_j} w_{i j} \cdot (y_i - <y_i>_i ) \cdot ( y_j - <y_i>_i )}{( {\sum_i} {\sum_j}w_{i j}) \cdot {\sum_i} ( y_i -  <y_i>_i )^2} $ where $y_i$ are the variable of interest (in this case $y_i = W_i$), $N$ is the number of pairs of counties separated by $d$ km where $d \in ( 10 \cdot n,10 \cdot ( n + 1 ) ]$, $w_{i j}$ are the weights, which we set to be $w_{ij} \sim {area}_i$ so that Moran's Index will reflect the spatial localization one sees while looking at the map.}
\label{fig1}       
\end{figure*}
\begin{figure*}
\resizebox{1\columnwidth}{!}{
  \includegraphics{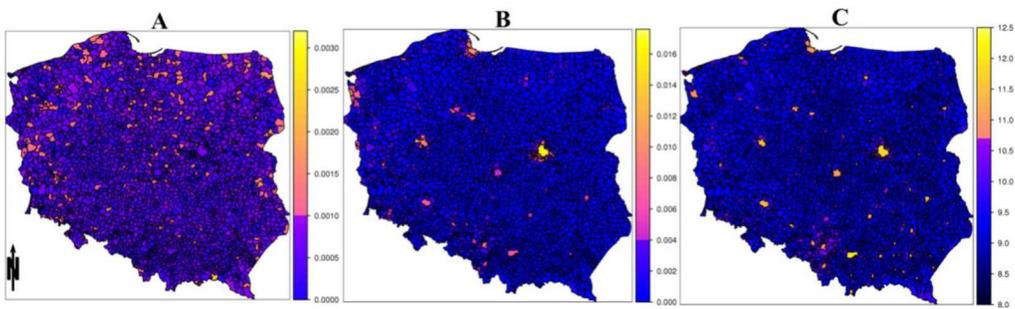}
}
\caption{\textbf{Maps Comparing Education to Economic Activity before and after liberalization }
\textbf{ \ref{fig2}A} The number of enterprises per capita in each county in 1989. 
\textbf{ \ref{fig2}B} The number of enterprises per capita in each county in 1994. 
\textbf{ \ref{fig2}C} The years of education per capita in various counties in 1988. 
\textbf{ \ref{fig2}A} shows the spatial distribution of number of the enterprises per capita in the year preceding the economic transition. This initial distribution is very close to a uniform random (Poisson) distribution and one does not observe any interesting spatial pattern. 
\textbf{\ref{fig2}B} The transition resulted in a marked change of the spatial pattern of economic activity: the economic activity after the transition is concentrated around the singular growth centers which are correlated to a high degree with the spatial patterns of the education levels as measured \textit{before} the transition (\textbf{ \ref{fig2}C}). }
\label{fig2}
\end{figure*}

%

\begin{figure}
\rotatebox{0}{\resizebox{1\columnwidth}{!}{%
  \includegraphics{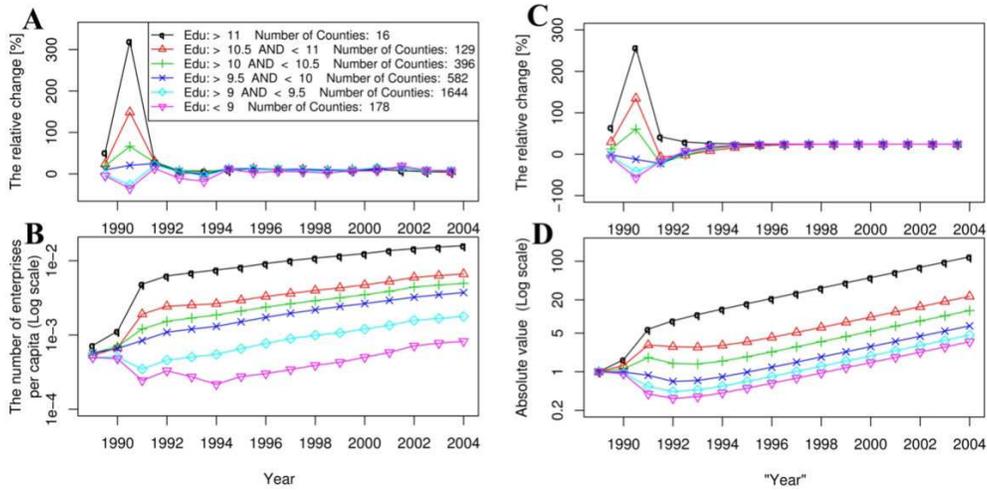}
}}
\caption{ 
\textbf{The time evolution of the number of enterprises per capita in various counties (aggregated  by their education level). }
\textbf{\ref{fig3}A,B }are empirical data while \textbf{\ref{fig3}C,D} are model predictions. 
\textbf{\ref{fig3}A.} The time evolution of the counties growth rates. The data are aggregated according to the average education level in intervals of 0.5 education years per capita. The first point, corresponding to the growth between 1989 and 1990 is largely representative for the communist regime since the Balcerowicz reform was introduced in 1990. Then, immediately after the liberalization the growth rates of the different counties diverged strongly: in the growth centers, the economic activity more then tripled while in most other counties it halved. Later on, the growth rates of all the counties became similar. 
\textbf{\ref{fig3}B} Nevertheless, the inequality between the growth centers and the rest of the counties continued to increase exponentially (the scale in this panel is Lin-log; see \textbf{\ref{fig5}} for a Lin-Lin graph).  
\textbf{\ref{fig3}C} and \textbf{\ref{fig3}D} represent the same effects as generated by a simulation based on the system \eqref{eqEIG}. where the monthly endogenous growth rates $a_j$'s corresponding to the above 6 groups were taken: -0.15, -0.1, -0.05, 0, 0.025 and 0.05. The $a_{ij}$'s were all set to 0.4.  One sees that \eqref{eqEIG} produces naturally the stylized facts in the data (\textbf{\ref{fig3}A} and \textbf{\ref{fig3}B}): initial divergence of growth rates $\frac{dW_i}{dt}$, followed by the equalization of the growth rates. One sees that the economic inequality between various counties $W_i - W_j$ continued to increase exponentially.}
\label{fig3}       
\end{figure}

\begin{figure}
\rotatebox{0}{\resizebox{0.89\columnwidth}{!}{%
  \includegraphics{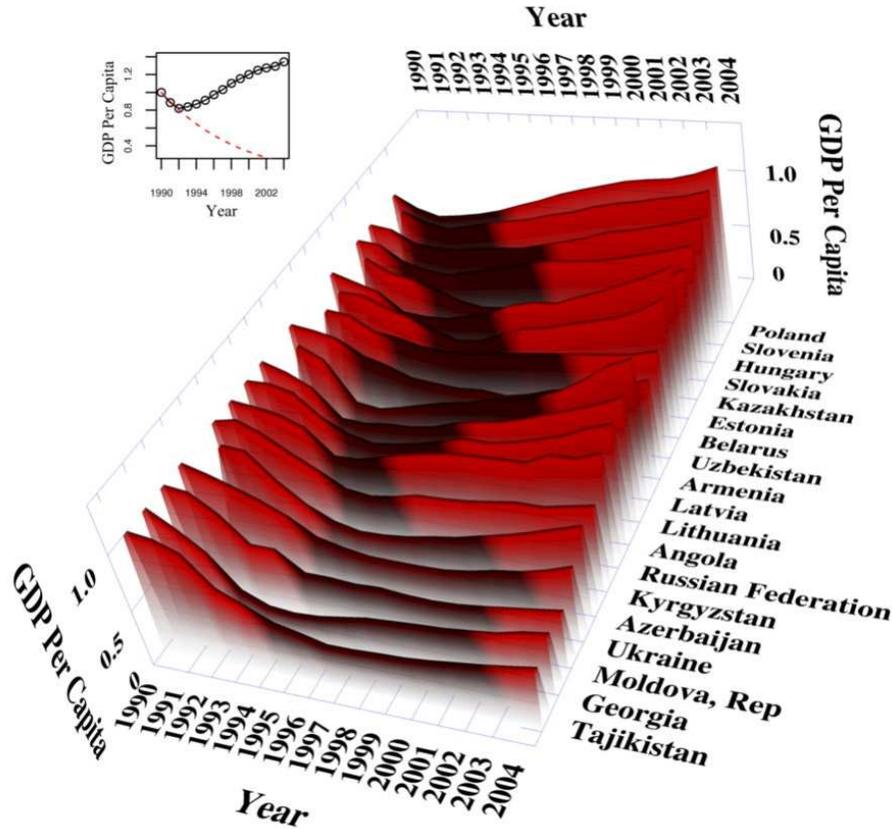}
}}
\caption{ \textbf{The "J-curve" evolution of the Eastern European GDP's after the liberalization.} Immediately after the economic liberalization, the economies of the Eastern European countries experienced strong decay followed by a growth period resulting in a pattern similar to a "J-curve" (here we use the term "J-curve" in a generic sense rather then the original meaning \cite{Brada_King1992} that was restricted to the balance of trade following a devaluation). Note that the details of this curve such as the magnitude of the initial decay and the time and rate of recovery varied between the countries. The inset shows the data from Poland and the decaying logistic fit extrapolated from the initial years of the decay. The marked departure (exponential growth) of the Polish economic activity from the extrapolated curve (exponential decay) indicates that the classical global logistic framework cannot explain the observed pattern. The "Microscopic Representation" version of the logistic dynamics on the other hand predicts this ubiquitous behavior by a generic mechanism.}
\label{fig4}       
\end{figure}

\begin{figure}
\resizebox{1\columnwidth}{!}{%
  \includegraphics{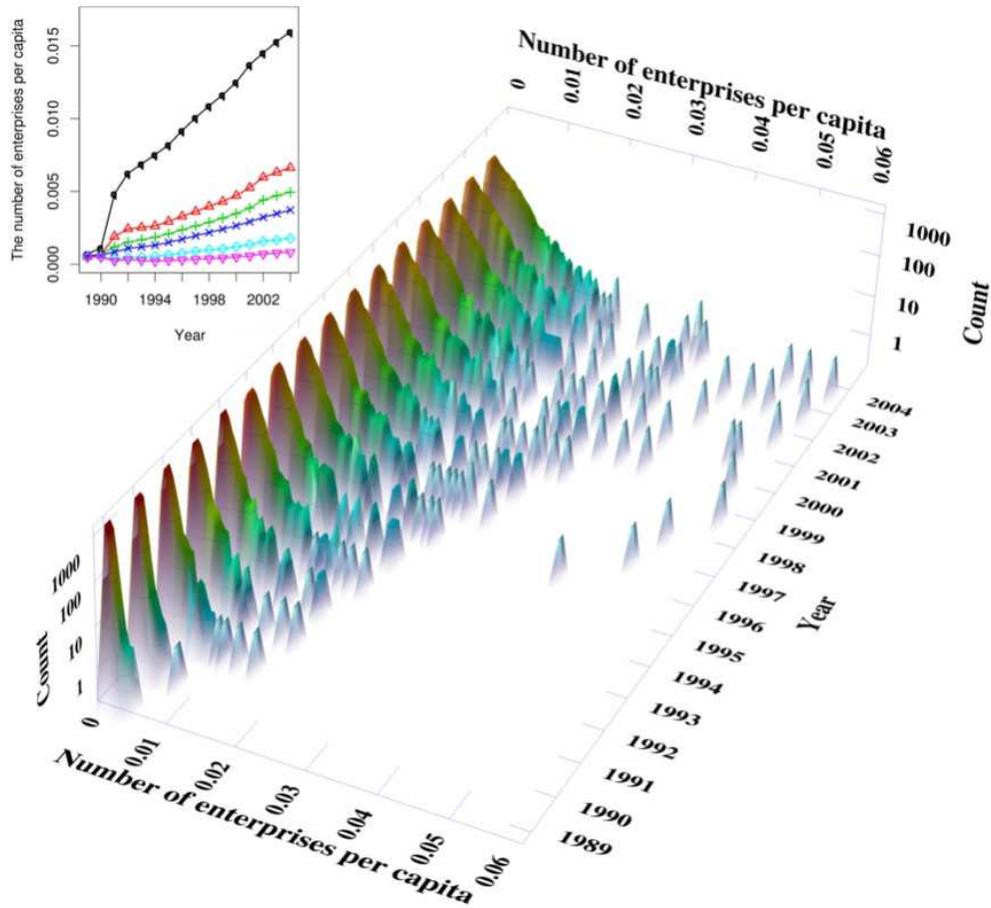}
}
\caption{The histogram (Lin-log plot) demonstrates the broadening distribution of enterprises per capita and the development of a fat tail as predicted by the microscopic model.\cite{Levy_ea2000} 
The inset shows the divergence in time of the counties economic activity per capita (the aggregation intervals are as in Fig \ref{fig3}).}
\label{fig5}       
\end{figure}

\begin{figure}
\rotatebox{270}{\resizebox{0.65\columnwidth}{!}{%
  \includegraphics{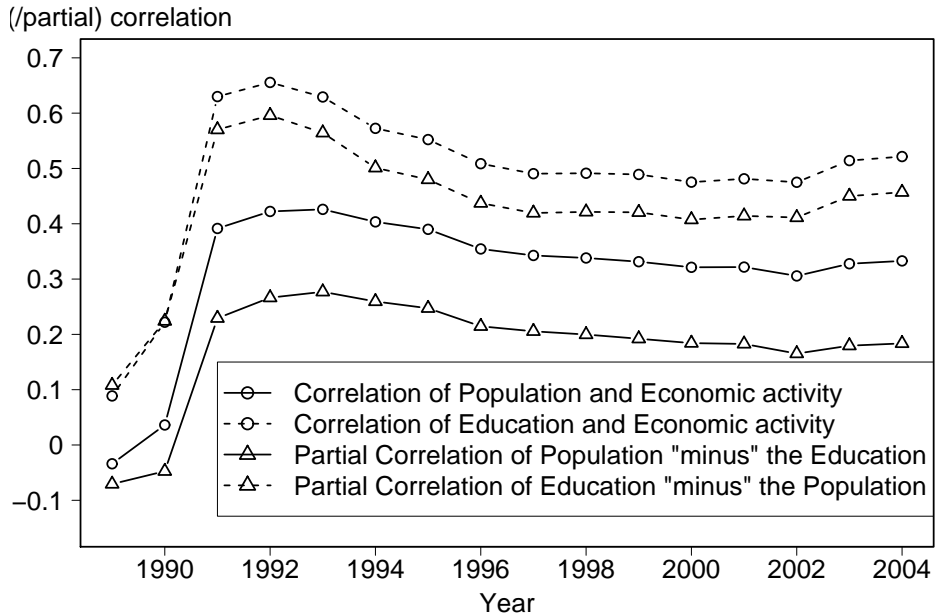}
}}
\caption{Correlations of the Number of
enterprises per capita to the education level and to the
population density respectively. 
 Before liberalization there was no significant influence of the 
education level on the economic activity (measured by the number of enterprizes per capita).
After liberalization one registered a dramatic increase in the 
correlation between those quantities.
Since the education level is correlated (0.357) to the population density,
we partialized out the correaltions of each of them to the economic activity.
The result is that the relevant quantity is the education level rather then the 
population density. 
 One observes a drop in the correlation in the subsequent years. 
This drop is a consequence of the diffusion: as time passes, the 
economic activity diffuses to counties that do not necessarily, have high 
education level.}
\label{fig6}       
\end{figure}

\bibliographystyle{epj}
\bibliography{polandepj}
%
%
%
\end{document}